# Towards Capacitive In-Memory-computing: A perspective on the future of AI hardware

Kapil Bhardwaj, Ella Paasio and Sayani Majumdar*

Faculty of Information Technology and Communication Sciences, Tampere University, FI – 33720, Tampere, Finland

*Email: sayani.majumdar@tuni.fi

**Abstract**

The quest for energy-efficient, scalable neuromorphic computing has elevated compute-in-memory (CIM) architectures to the forefront of hardware innovation. While memristive memories such as resistive random-access memories (RRAMs), phase-change memory (PCM), magneto resistive random-access memory (MRAM), ferroelectric random-access memories (FeRAM) have been extensively explored for synaptic implementation in CIM architectures, their inherent limitations, including static power dissipation, sneak-path currents, and interconnect voltage drops, pose significant challenges for large-scale deployment, particularly at advanced technology nodes. In contrast, capacitive memories offer a compelling alternative by enabling charge-domain computation with virtually zero static power loss, intrinsic immunity to sneak paths, and simplified selector-less crossbar operation, while offering superior compatibility with 3D Back-end-of-Line (BEOL) integration. This perspective highlights the architectural and device-level advantages of emerging non-volatile capacitive synapses, including metal–ferroelectric–metal (MFM), metal–ferroelectric–semiconductor (MFS), ferroelectric field-effect transistors (FeFETs), and hybrid configurations. We examine how material engineering and interface control can modulate synaptic behavior, capacitive memory window (CMW), and multilevel analog storage potential. Furthermore, we explore critical system-level trade-offs involving device-to-device variation, charge transfer noise, dynamic range, and effective analog resolution. Capacitive memories, we argue with custom-built stacks, have the potential to become a foundational technology for the next generation of extremely energy-efficient neuromorphic computing platforms.

## 1. Introduction

The rapid expansion of data-centric applications, from real-time edge analytics to autonomous systems, has accelerated the demand for new classes of hardware capable of delivering high energy efficiency, low latency, and scalable performance [1]. Conventional von Neumann architectures, constrained by the memory-wall bottleneck and the energy-intensive data movements between processing and memory units, are increasingly inadequate for supporting modern artificial intelligence (AI) workloads [2]. Neuromorphic computing, inspired by biological neural systems, offers an alternative paradigm by tightly integrating memory and computation through distributed, parallel, and event-driven architectures [3].

To realize the full potential of neuromorphic computing, the compute-in-memory (CIM) paradigm has emerged as a pivotal architectural innovation [4]. In CIM, critical operations such as matrix-vector multiplications (MVMs) are executed directly within memory arrays, eliminating the frequent and costly data movements between separate processing and storage units. Analog or mixed-signal CIM implementations leverage the inherent parallelism of crossbar structures, where input vectors are applied as voltages along wordlines (WLs) and accumulated current signals are collected from bitlines (BLs). The multiply operation follows Ohm's law, with each crosspoint's conductance acting as the multiplying factor, while accumulation along BLs occurs by Kirchhoff's current law, making multiply-and-accumulate (MAC) operations extremely fast and energy-efficient.

Among the emerging memory technologies explored for CIM applications, resistive memories, such as oxide filamentary resistive random-access memory (RRAM) [5], have attained significant attention due to their simple structure and non-volatile switching properties. Following the experimental realization of a memristor in 2008 using a $TiO_2$-based solid-state device [6], research efforts intensified in both material innovations and mathematical modelling to better capture and exploit memristive behaviour [7-10]. Other emerging resistive memory technologies, such as phase-change memory (PCM) [11], spin-transfer torque magnetoresistive RAM (STT-MRAM) [12], spin-orbit torque magnetic RAM (SOT-MRAM) [13], ferroelectric RAM (FRAM) [14], ferroelectric tunnel junctions (FTJs) [15-17], and ferroelectric field-effect transistors (FeFETs) [18-19], have also demonstrated promising switching characteristics, retention, and endurance behaviour, offering potential pathways toward scalable CIM solutions [20].

In addition to non-volatility, these emerging technologies offer advantages such as multi-bit programmability and compatibility with advanced CMOS nodes, positioning them as strong contenders for embedded memory and neuromorphic computing platforms. Early prototypes leveraging these technologies have demonstrated the feasibility of CIM architectures, particularly for analog MVMs critical to neural network inference tasks [19, 21-22, 23].

Beyond metal-insulator-metal (MIM) memristive devices, CMOS-based components, such as operational transconductance amplifiers (OTAs), have also been used to emulate first-order memristive dynamics [24-25]. These emulator circuits have found applications in neuromorphic sensing [26] and CMOS-compatible emulator circuit designs [27], offering a bridge between solid-state device physics and scalable circuit-level implementations.

Over the past decade, resistive memory-based crossbar architectures have emerged as strong candidates for enabling energy-efficient CIM systems, particularly in neuromorphic computing contexts. The ability to store analog weights directly in memory arrays is critical for performing MVM operations in situ, reducing data movement and improving overall energy efficiency. In this regard, RRAM based systems have shown major progress. For example, the 48-core NeuRRAM chip, integrating around 3 million RRAM devices monolithically on CMOS, achieved inference accuracies comparable to 4-bit quantized digital models, reporting ~0.98% test error on MNIST and 14.3% on CIFAR-10, by co-optimizing device programming and network training [28]. Fully memristive convolutional neural networks (CNNs) have demonstrated over 96% MNIST accuracy, compensating for device-level imperfections through hybrid training schemes [29].

Advances in device-to-device (D2D) uniformity have further propelled RRAM technology toward supporting high-precision analog computing. A 64×64 passive RRAM crossbar achieved grayscale image encoding with less than 4% average error, reaching effective precision levels of ~6–7 bits per cell [30]. Notably, a recent study has reported 2,048 discrete conductance levels in 256×256 arrays with one transistor-one memristor (1T1R) weight elements fabricated in commercial foundries, marking a significant milestone in analog resolution for RRAM-based CIM systems [31]. Furthermore, RRAM-based CIM architectures have demonstrated MAC energy consumptions as low as ~190 fJ/MAC [32], with constant-current crossbar techniques (C3CIM) reducing this further to ~30 fJ/MAC for 1-bit MACs [33]. Larger prototypes, such as a 28 nm implementation with a 576k RRAM-based CIM macro, achieved energy efficiencies of ~35.6 TOPS/W, equivalent to ~$10^{-14}$ J per operation [34]. In standalone neural network implementations, memristor-CNN systems have demonstrated over 100× energy savings compared to conventional GPUs [29], highlighting the inherent advantage of analog parallelism in memristive crossbar arrays.

While memristive CIM accelerators have made significant advances, persistent challenges remain. Issues such as inherent randomness-driven cycle-to-cycle variation [35], significant IR drops in low-resistance states [36], and sneak-path currents [37] continue to limit scalability. Furthermore, the need for selector integration complicates back-end-of-line (BEOL) stacking processes [38-39], constraining large-scale integration. These limitations have motivated the exploration of alternative memory technologies better suited for robust and scalable neuromorphic computing.

Capacitive memories, particularly those based on ferroelectric capacitors (FeCAPs), have emerged as promising candidates for next-generation CIM architectures [40]. By encoding information in programmable capacitances rather than resistance states, capacitive CIM adopts a fundamentally different computation model. Here, computation is driven by displacement currents rather than conductive paths, inherently eliminating issues like static leakage, sneak-path effects, and IR drop. This facilitates true selector-less crossbars with simpler and more scalable peripheral circuitry, paving the way for energy-efficient, high-density neuromorphic processors.

Although memcapacitors were proposed long ago [41], early memcapacitive networks did not gain significant attention primarily due to the lack of non-volatile data retention and the absence of analog programmability. However, the emergence of CMOS BEOL-compatible binary ferroelectric oxides, such as hafnium-zirconium oxide (HZO) [42-44], has reignited interest in ferroelectric-based non-volatile capacitive devices. FeCAPs implemented with HZO exhibit non-destructive, low-voltage readout, high endurance (>$10^{11}$ cycles), long-term data retention (>10 years) [45], and sub-nanosecond switching speeds [46], meeting critical energy and reliability benchmarks required for neuromorphic computing. Their extremely high off-state impedance and compatibility with low-voltage sensing (<1.5 V) make them ideal for energy-constrained edge AI applications. Moreover, their two-terminal structure enables dense, vertically stacked 3D arrays fabricated using BEOL processes, offering a pathway toward highly scalable neuromorphic accelerators [47].

Overall, capacitive memories offer a fundamentally superior pathway for scalable, energy-efficient, and reliable neuromorphic hardware compared to resistive memories. By aligning better with the core requirements of low-power, high-density, and high-accuracy in-memory computing, capacitive CIM represents a pivotal advancement in the evolution of neuromorphic accelerators.

In this perspective article, we present a comprehensive, top-down analysis of capacitive memory-based CIM architectures for neuromorphic AI hardware. Section 2 provides a comprehensive overview of the various memristive memories such as RRAM, PCM, MRAM and ferroelectric resistive memories. Section 3 describes the inherent operating-level advantages and mechanism of the capacitive memories in CIM crossbars. Section 4 delves into the device physics underpinning ferroelectric capacitors, including polarization dynamics, charge shielding, and interface engineering to achieve optimal performance parameters. Section 5 discusses design trade-offs across device, circuit, and system layers, highlighting challenges and opportunities for achieving scalable, energy-efficient charge-domain computing through cross-layer co-optimization in capacitive CIM architectures. Finally, Section 6 outlines an outlook for capacitive CIM as a foundational technology for future AI accelerators.

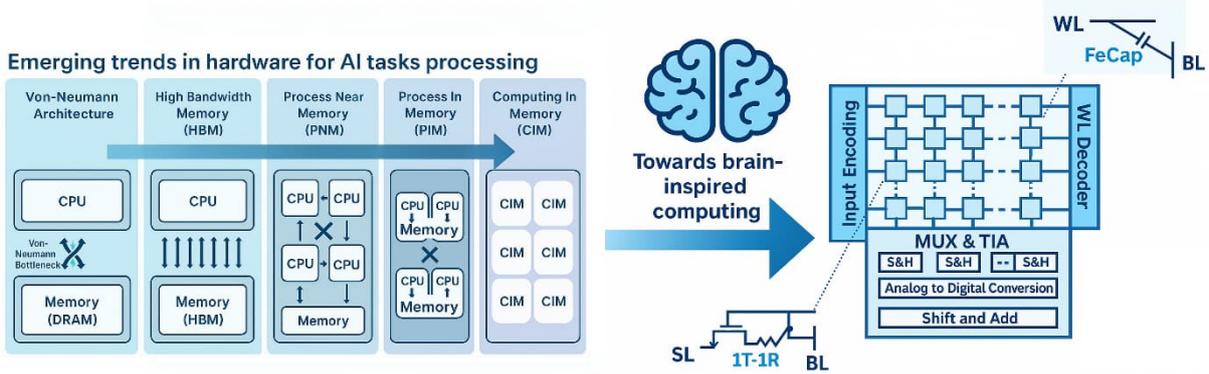

**Figure 1. Evolution of Hardware implementation of AI computation:** Emerging hardware trends for AI acceleration, highlighting the shift from traditional von Neumann architectures to computing-in-memory (CIM) systems. FeCAP-based crossbar arrays (right) enable brain-inspired, highly parallel charge-domain MAC operations with reduced leakage, eliminating sneak-path currents compared to resistive crossbar arrays (left) showing promising routes for scalable, energy-efficient neuromorphic processors.

## 2. Resistive Memories: Enabling low-power AI hardware

Figure 1 shows evolution of hardware for AI computation, starting from traditional von Neumann architecture to processing near and in-memory systems. In memristive CIM architectures, each synaptic weight is physically realized by programming a two-terminal resistive device into a specific conductance state, typically modulated by controlling electronic, ionic, or spin transport mechanisms within the material. Computation in these systems relies on analog accumulation of currents. Input voltages are applied along WLs, and the resultant weighted summation of the currents (conductance matrix x input vector) is collected at the BLs. Each crosspoint effectively performs a multiply operation via Ohm's law, and the resultant currents are accumulated along BLs following Kirchhoff's current law.

Typically, memristive devices exhibit two primary programmable states: a low-resistance state (LRS) and a high-resistance state (HRS), as depicted in the I-V characteristic curve (Figure. 2a). For RRAM, PCRAM and MRAM cells, the LRS typically lies in the range of ~1–10 k$\Omega$, while the HRS spans ~100 k$\Omega$–1 M$\Omega$. Although the HRS minimizes conduction, non-negligible leakage currents persist under modest read voltages, leading to static power consumption even across idle memory cells. Ferroelectric devices, particularly in scaled nodes, exhibit inherently high resistivity [16]; however, their non-linear I-V behavior, especially the variable slope in the low-voltage regime, complicates precise analog weight encoding and accumulates inaccuracies in large-scale deployments, where thousands of parallel operations magnify device-level variability.

While significant progress has been made in memristive CIM accelerators, challenges remain in reducing dynamic energy overheads and maintaining fidelity as array sizes scale [28]. In a typical crossbar architecture (Figure. 2b), each memristive device at the BL-WL crosspoint modulates its conductance to encode a synaptic weight, with the total bitline current representing the accumulated sum. Although conceptually simple (Figure. 2c), crossbar arrays are highly sensitive to device non-idealities and interconnect parasitics. Intrinsic stochasticity, particularly prevalent in filamentary conduction-based devices, remains a critical bottleneck affecting stability and reproducibility.

*Resistive Random-Access Memory (RRAM)*

RRAM devices typically employ a simple MIM structure, where an insulating oxide layer, such as $HfO_2$, $TiO_2$, or $Ta_2O_5$, is sandwiched between two metal electrodes [6]. The conduction mechanism is primarily governed by the formation and dissolution of nanoscale conductive filaments within the dielectric. Under an applied electric field, mobile ions (e.g., oxygen vacancies) migrate, inducing localized redox reactions that generate conductive filaments composed of reduced oxide phases or metallic inclusions. In the LRS, a filament bridges the electrodes, enabling conduction through Ohmic transport or trap-assisted tunneling, depending on filament composition and morphology. In the HRS, the rupture or constriction of the filament suppresses carrier transport, with conduction primarily governed by mechanisms such as Poole–Frenkel emission, Schottky emission, or space-charge-limited conduction (SCLC), depending on material and interface properties.

While RRAM-based crossbar arrays demonstrate promising performance in moderate sizes, scaling to larger arrays faces critical hurdles. The kΩ-range LRS induces significant IR drops along the interconnects [36], degrading linearity and compromising computation accuracy [48]. Furthermore, read disturb effects, where repeated sensing gradually alters the programmed conductance states, impose endurance constraints, making RRAM less suitable for continuous on-chip training. Achieving stable memristive weights over billions of switching cycles remains non-trivial.

Sneak-path currents, a pervasive issue in passive crossbars, further complicate scaling (Figure. 2d). Although self-rectifying RRAM devices with inherent rectification ratios >$10^4$ have been demonstrated [49], large arrays (>256×256) still necessitate selector integration. However, adding access devices such as transistors or selectors increases the footprint beyond the ideal 4F², and complicates BEOL integration [38-39], posing a major barrier for high-density 3D stacking efforts.

In terms of integration density, recent architectures like NeuRRAM have demonstrated feasibility, occupying only a few tens of mm² in 130 nm technology [28], yet achieving brain-scale synaptic densities requires orders of magnitude higher integration, approaching billions of synapses on-chip. This drives research towards restoring the 4F² cell size either via selector-less designs or advanced 3D stacking techniques. Multi-layered memristive crossbars, stacking memory layers vertically, offer a promising path to scaling; however, controlling device variability and IR drop propagation becomes increasingly challenging in 3D configurations, particularly for filamentary memristive devices.

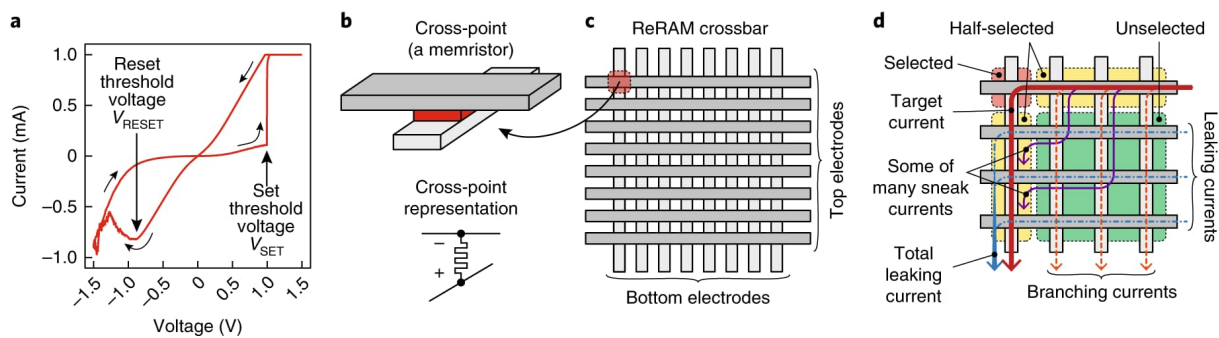

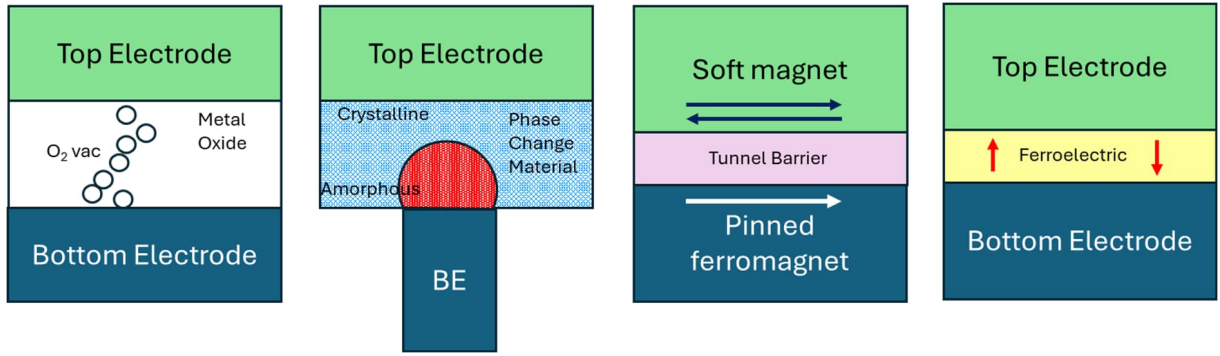

**Figure 2.** (Top panel) RRAM-Based Crossbar Arrays: (a) Typical I–V characteristics showing SET and RESET thresholds; (b–c) Cross-point structure and array architecture; (d) Sneak-path currents through half-selected and unselected cells degrade read accuracy and energy efficiency in large-scale resistive memory systems [ Figure reproduced with permission from [48]]. (Bottom panel) Resistive switching mechanisms of oxide RRAM, PCM, MRAM and FRAM devices.

*Phase-Change Memory (PCM)*

PCM operates on the principle of reversible phase transitions between amorphous and crystalline states within a phase-change material, typically a Ge–Sb–Te (GST) alloy. These two states exhibit a significant contrast in electrical resistance (Figure. 2), and by applying electrical pulses, the material can be switched between them, using the resistance change to represent data. PCMs are two-terminal memristive devices that usually require large currents for the RESET process, melting the phase-change material into the amorphous phase, which negatively impacts their energy efficiency.

Various studies have focused on reducing operation currents by minimizing device dimensions, but aggressive scaling increases fabrication costs, and further reductions in RESET current remain limited. The key requirements for PCM devices for electrical data storage include high endurance, low RESET current, fast SET speed (≤100 ns), high retention (typically 10 years at 85 °C, though different requirements exist for embedded memories), good scalability (<45 nm node), and low intra- and inter-cell variability. While these requirements have been achieved in individual PCM devices, demonstrating >$10^{12}$ endurance cycles, <10 µA RESET current, ~25 ns SET speed, projected 10-year retention at 210 °C, and sub-20 nm scalability [50], replicating such performance uniformly in large arrays remains challenging.

A thorough review of PCM memories and CIM accelerators is available in the literature [51] and is beyond the scope of this work. Briefly, PCM technology has matured over many years, and by leveraging specially designed architectures, high computing accuracy has been achieved in PCM-based CIM accelerators. In a recent work, IBM researchers have demonstrated cross-point arrays built with phase-change tunable memresistors is capable of implementation of computationally efficient speech recognition and transcription tasks [52]. Their analog-AI chip, integrating 35 million PCM devices across 34 tiles, supports massively parallel inter-tile communication and achieves up to 12.4 tera-operations per second per watt (TOPS/W) sustained performance. It demonstrated near software-equivalent (SWeq) accuracy for a small keyword-spotting network and near-SWeq accuracy on the much larger MLPerf8 recurrent neural network transducer, mapping over 45 million weights onto more than 140 million PCM devices across five chips.

However, PCM technology still faces known drawbacks, including resistance drift, and high on-state currents, requiring specially designed architectures to support high-precision analog computing [53].

*Magnetic Random-Access Memory (MRAM)*

MRAM is a current-controlled technology based on magnetic tunnel junctions (MTJs), where two ferromagnetic (FM) layers are separated by a thin insulating barrier [54]. One FM layer (pinned layer) has a fixed magnetic orientation, while the other (free layer) can switch its orientation. In CMOS processes, MTJs are embedded between two metal layers. When the magnetizations of the two FM layers are parallel, the device exhibits low resistance ($R_L$), and when antiparallel, it exhibits high resistance ($R_H$). These resistance differences are used to represent binary data, although the resistance levels are generally lower than those in other NVM technologies.

Commercial MRAM technologies are categorized into spin-transfer torque (STT-MRAM) and spin-orbit torque (SOT-MRAM) types. STT-MRAM operates via spin-transfer torque, while SOT-MRAM utilizes spin-orbit torque mechanisms [55]. The critical parameters for evaluating STT-MRAM cells include resistance, tunneling magnetoresistance (TMR) ratio, switching current, and thermal stability. A key challenge for STT-MRAM is the simultaneous achievement of low switching currents and high thermal stability. SOT-MRAM offers advantages, being four times faster and more reliable in terms of read disturbance and dielectric breakdown compared to STT-MRAMs. At the architectural level, SOT-MRAM outperforms STT-MRAM in read/write energy and latency, although it incurs a slightly larger footprint due to the 2-transistor, 1-MTJ (2T-1MTJ) architecture [56].

Crossbar arrays of MRAM devices have been demonstrated for CIM. However, a major challenge in scaling STT-MRAM crossbars is the low resistance of MRAM cells, which leads to significant power consumption. Recently, Samsung Electronics demonstrated a 64×64 MRAM crossbar array that overcomes this challenge by employing resistance summation techniques for analog MAC operations [57]. Integrated with 28 nm CMOS readout electronics, this array enabled the implementation of a two-layer perceptron for classifying 10,000 MNIST digits with 93.23% accuracy. Further emulations of deeper networks, such as an 8-layer Visual Geometry Group-8 (VGG-8) neural network, achieved 98.86% classification accuracy, close to software baseline values of 95.24% and 99.28%, respectively.

*Ferroelectric Resistive Memories*

Ferroelectric devices work on the mechanism of voltage-driven switching of spontaneous polarization i.e., internal electric dipole moments of the material. Dielectric materials with a non-centrosymmetric structure inherently exhibit piezoelectricity; however, a subset of these materials possesses a unique axis of symmetry, making them polar and leading to spontaneous electric polarization that persists even after the external electric field is withdrawn. This property makes ferroelectric materials particularly attractive for non-volatile data retention applications.

In conventional perovskite ferroelectrics, polarization typically has a displacive origin. In contrast, fluorite-structured materials such as doped-$HfO_2$ exhibit ferroelectricity through the formation of a non-centrosymmetric orthorhombic phase, induced by factors such as strain, surface and interface energy, and kinetic stabilization. The compatibility of these fluorite materials with CMOS processes, coupled with their scalability potential, has brought them to the forefront of non-volatile memory research over the past decade.

Fluorite-based ferroelectric devices encompass 1T1C FeRAMs, FeFETs, and ferroelectric tunnel junctions (FTJs). FeRAMs operate similarly to dynamic random-access memories (DRAMs) but use ferroelectric capacitors instead of conventional dielectric capacitors, enabling non-volatile charge storage. FeFETs are MOSFETs in which the gate oxide is replaced by a ferroelectric oxide, allowing the storage of multi-bit information by varying device conductance states based on the polarization state of the ferroelectric gate stack. FTJs, on the other hand, operate on the principle of quantum mechanical tunnelling, where polarization switching alters the tunneling barrier's height, and in some cases, its width, resulting in ON/OFF switching behavior.

Ferroelectric devices offer several advantages, including sub-100 ns switching speeds [58], forming-free switching, low device currents, and excellent endurance and reliability. However, challenges such as ultra-low current levels, device-to-device (D2D) and cycle-to-cycle (C2C) variability in scaled devices [16] have raised concerns. Recent experimental advances in thin-film ferroelectric technology have addressed these limitations, leading to the development of low-leakage, high-endurance ferroelectric thin-film capacitors with recoverable fatigue [59] and wafer scale uniformity [60] and exhibit potential for 3D vertical stacking. Leakage current issues remained a challenge in these devices impacting device reliability over prolonged bias stressing, however, recent results [61, 62] showed that cooling the devices with liquid nitrogen even can provide significantly improved performances, improving device endurance and stability.

In addition to memory applications, FeFET-based CIM operations have also been demonstrated [63, 64]. By combining ternary content-addressable memory (TCAM) and CIM arrays based on FeFETs, machine learning models with the capacity to learn new classes of data have been implemented. These hardware systems achieve significantly reduced training overheads, attaining 95.14% accuracy in few-shot learning tasks. The dual role of FeFETs as compact logic and storage elements enables dense integration of CIM and TCAM structures, driving highly efficient neuromorphic systems [63]. Further, Dutta et al. demonstrated "lifelong learning" capabilities with BEOL-integrated TCAM arrays using FeFET-based hardware [65]. This study further contributed by introducing the FeMFET memory concept, where a TiN/HZO/TiN FeCap stack is integrated into the BEOL of CMOS gate stacks, enabling MAC operations with good reproducibility and minimal current deviation.

Beyond FeFETs, significant progress has been made in FeCAP-based CIM architectures. FeCAPs offer a considerable capacitive memory window (CMW), which is critical for analog and multilevel storage applications. CMW is associated with asymmetry in C-V curve (Figure. 3) and measured as difference in capacitances at zero bias or very low read bias within a C-V loop of a FeCap. Recent studies have demonstrated that engineering asymmetric ferroelectric stacks via work function asymmetry [66], interfacial layer optimization [67], introducing charge-shielding layers [68], and employing co-design approaches with other devices [69] can significantly enhance CMW, enabling reliable multilevel programming with minimal state overlap.

Recent benchmarking studies have shown that FeCAP-based capacitive CIM crossbars deliver 31–48% lower total inference energy compared to RRAM- and STT-MRAM-based arrays across standard deep neural network architectures such as VGG-8, ResNet-18, and ResNet-50 [70]. Moreover, FeCAP arrays exhibit negligible performance degradation even after extensive cycling and repeated read operations, underscoring their exceptional suitability for inference-dominated workloads. These results demonstrate that charge-domain CIM not only improves energy efficiency but also enhances analog computation fidelity by decoupling computation from resistive interconnect losses. Capacitive readout schemes provide fine-grained control over resolution, dynamic range, and noise resilience through careful sizing of reference capacitors and optimization of sensing circuits, presenting new opportunities for advanced device-circuit co-design in neuromorphic computing systems.

## 3. Capacitive Memories: A Leap Ahead of Resistive Counterparts

Due to their unique features, particularly CMOS BEOL compatibility, scalability, and low-power operation, FeCAPs have emerged as the most promising candidates for realizing memcapacitors, i.e., capacitive non-volatile memories. The programmability of FeCAPs arises from switching between two polarization states (P↑ and P↓), which directly modulates the dielectric constant (and thus capacitance) of the device, as shown in Figure 3. FeCap stack when subjected to electric fields switches the polarization state, allowing clear separation between high-capacitance (HCS) and low-capacitance (LCS) states. The capacitance window (ΔC = HCS - LCS) is crucial for analog precision. A larger ΔC enables better multi-bit resolution with lower overlap between states, improving inference accuracy. Moreover, because FeCAP switching involves polarization dynamics rather than filament formation, the process is highly reproducible, low-energy (~pJ per switching event), and highly endurance-capable (>$10^{11}$ cycles), as demonstrated in recent HZO-based devices [45-46, 70]

In FeCAP-based CIM crossbars (schematically shown in Figure 4), each memory cell consists of a programmable ferroelectric capacitor. The synaptic weights are stored as discrete capacitance states, which correspond to different polarization configurations within the ferroelectric material.

The charge-domain computation paradigm enabled by FeCAP-based CIM architectures offers several intrinsic advantages that collectively address the fundamental bottlenecks faced by resistive memory-based systems. Since FeCAPs act as ideal capacitors during the hold phase, no steady-state conductive path exists, thereby eliminating leakage currents and static power dissipation even in large crossbar arrays. The absence of conduction not only reduces energy loss but also provides intrinsic immunity to sneak-path currents, enabling true 0T-1C crossbar architectures without requiring additional selector devices or access transistors. Furthermore, the extremely small displacement currents involved in charge transfer operations, typically in the picoampere to nanoampere range, ensure that IR voltage drops across interconnects remain negligible, preserving the input-output linearity even as array sizes scale up.

The FeCAP operation cycle consists of two carefully orchestrated phases as described in Figure 4c [69]. **Phase Φ₁: Initialization (Pre-charge Phase)**-In this phase, an initialization voltage is applied along the WL to establish a known charge distribution across the FeCAPs. Depending on the stored polarization state, the effective capacitance value determines the amount of displacement charge available for subsequent computation. **Phase Φ₂: Charge Transfer (Accumulation Phase)**-After initialization, the array enters the accumulation phase, where input signals (small voltage perturbations) are applied across the FeCAPs. The stored polarization state modulates the transferred displacement charge, which is accumulated onto dedicated reference capacitors ($C_{ref}$) or sensing nodes. The final output voltage at each bitline is proportional to the sum of input voltages weighted by the programmed capacitances as;

$$V_j = \frac{1}{C_{ref}} \sum_i C_{FE(i,j)} \, X \, V_{WL(i)} \qquad (1)$$

Another critical benefit of FeCAP-based CIM is the ability to perform non-destructive read outs. Since the readout is based on small-signal perturbations, the stored polarization state of the FeCAP remains intact across multiple read cycles, ensuring superior endurance and long-term data retention.

In addition, charge-domain architectures offer fine-grained analog control, e.g., the output resolution, dynamic range, and sensing characteristics can be precisely tuned by adjusting the $C_{ref}$ and the read bias conditions during the charge accumulation phase, allowing system designers to flexibly trade-off between energy, precision, and speed based on application requirements. Compared to current-domain summation, charge-domain operations are inherently more resilient to noise accumulation, owing to the voltage-based nature of signal integration across capacitors rather than through resistive paths. Finally, FeCAP-based CIM platforms demonstrate exceptional energy efficiency, with recent system-level studies reporting 30–50% lower inference energy compared to resistive CIM approaches, a decisive advantage particularly for ultra-low-power edge AI applications [70].

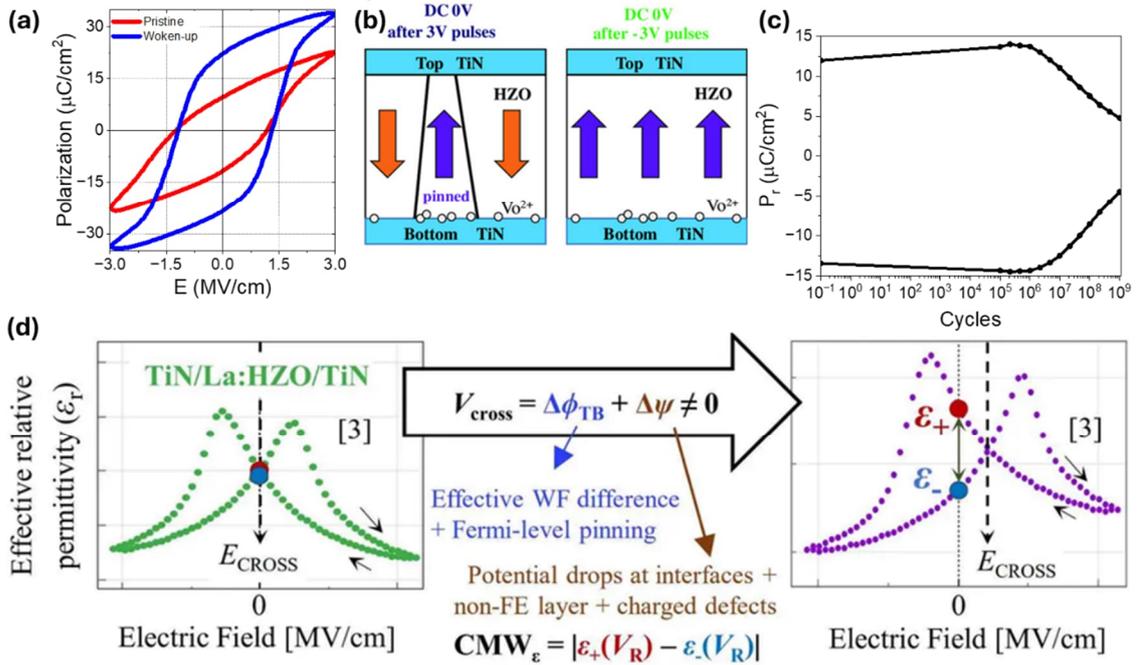

**Figure 3.** Polarization switching characteristics of HZO Capacitors: (a) Measured P–V characteristics showing a distinct memory window between up and down polarization states within the electric field range of +3 MV/cm to -3MV/cm giving rise to high-capacitance (HCS) and low-capacitance (LCS) states in the device. (b) Polarization-dependent oxygen vacancy alignment explaining the modulation of relative permittivity ($\varepsilon_r$) based on pulse history and DC bias. (c) Endurance characteristics of the capacitors showing setting-in of fatigue after $10^7$ switching cycles, however, no hard breakdown until $10^9$ cycles. (d) $\varepsilon_r$ vs. V plots for a symmetric capacitor showing no CMW while adding effects like potential drops at interfaces and Fermi-level pinning led to asymmetry in a $\varepsilon_r$ - V loop opening a CMW. Figures reproduced with permission from [59, 60].

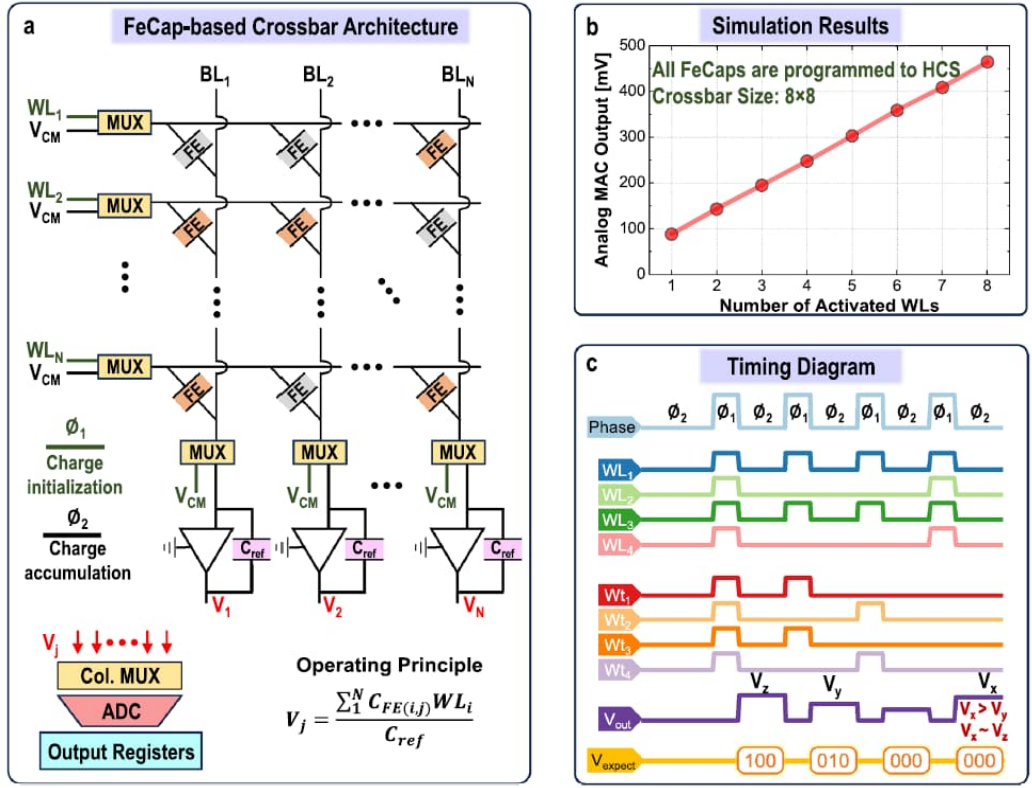

**Figure 4.** FeCap-Based Charge-Domain CIM Operation: (a) Crossbar architecture using ferroelectric capacitors and charge reference integration for analog MAC; (b) Simulation results showing linear accumulation with increasing active wordlines; (c) Timing diagram illustrating two-phase ($\Phi_1$, $\Phi_2$) charge and transfer operations enabling accurate vector-matrix multiplication. [Figures reproduced with permission from [69]]

Table 1 highlights key architectural and device-level differences between Memristive memories-based and FeCAP-based CIMs. Capacitive CIM offers advantages in static power, sneak path immunity, and analog linearity, making it better suited for energy-efficient and scalable neuromorphic hardware. Its compatibility with BEOL and selector-free operation further enhances 3D integration potential.

**Table 1.** Comparative Analysis of various synaptic memory candidates for CIM Architectures

| Parameter | RRAM-Based CIM | PCM-Based CIM | MRAM-Based CIM | FRAM-Based CIM | FeCAP-Based CIM (Ferroelectric Capacitor Crossbar) |
|---|---|---|---|---|---|
| **Static Power** | ~50–200 pJ per 128×128 MAC (inference) [68] | Negligible (few eV) [71] | No static power; leakage only through transistors [72] | Negligible static power [69] | ~3.8 pJ per MAC (128×128 array) [68] |
| **Sneak-Path Currents** | Significant; can exceed selected cell's current [37, 39] | Present without selectors [71] | Eliminated by 1T-1MTJ structure [80] | No sneak paths [69] | None; inherently selectorless [70] |
| **IR-Drop (Voltage Drop)** | ~7.7 mV across 256 cells at 1.7 V (~4.5% loss) [36] | Notable in large arrays [81] | Present during fully parallel read [80] | High inherent cell resistance; | Minimal due to negligible steady-state current and |
| | | | | negligible drop [69] | high impedance [76] |
| **Read Disturb** | Requires ~0.1–0.2 V for non-destructive read [73] | Negligible; non-destructive [81] | >$10^9$ disturb-free reads [80] | Non-destructive read at low voltages [69] | Non-destructive, near-zero small-signal read voltage [70] |
| **Endurance (Write Cycles)** | $10^5$–$10^7$ typically; up to $10^9$–$10^{12}$ in lab tests [73] | $10^6$–$10^8$ cycles [78] | Extremely high; >$10^{12}$–$10^{15}$ cycles [78] | $10^{10}$–$10^{15}$ cycles [78] | $10^{10}$–$10^{12}$ cycles for HZO-based FeCAP [69] |
| **Selector Requirements** | 1T1R or 1S1R configurations; self-rectifying cells partially mitigate [39] | Requires selector or 1T1R [83] | Inherent 1T-1MTJ structure [80] | Integrated selector via FeFET structure [69] | No separate selector required [69] |
| **3D Stacking** | Possible via self-rectifying arrays, at energy cost [49] | Feasible; demonstrated [82] | No commercial 3D MRAM yet; research ongoing [79] | Dual-layer 3D FeRAM (32 Gb) demonstrated [78] | BEOL-compatible; two-terminal structure enables vertical stacking [70] |
| **Small-Signal Read Voltage** | ~0.1 V [74] | ~0.1–0.3 V [50] | ~50–200 mV [72] | ~0.8–1.0 V (FeFET threshold sensing) [69] | Few millivolts; much lower than coercive voltage [70] |
| **Analog Linearity** | 4–6 bits per cell practical [75] | Nonlinear conductance; mitigated by cell averaging [50] | Limited by tunneling magnetoresistance (TMR); binary nature dominates [72] | Hysteresis and saturation affect analog linearity [69] | Linear charge accumulation; limited dynamic range (~2 bits) [69] |
| **Energy per MAC** | 100–200 fJ/MAC (130 nm CMOS) [28] | ~0.1 pJ/MAC (10–12 TOPS/W) [50] | ~0.04–0.1 pJ/MAC [72] | 885.4 TOPS/W for 2-bit MAC (~$1\times10^{-15}$ J per operation) [77] | Sub-10 fJ/MAC achievable; sub-femtojoule regime projected [69] |

## 4. Device-Level Physics and Engineering of Non-Volatile FeCAPs

Capacitive synapses are based upon asymmetry in capacitance – voltage (C-V) characteristics, which can be realized through numerous device architectures. Utilizing ferroelectric switching to induce memcapacitive behaviour is promising due to the voltage controlled write mechanism, CMOS compatibility and scalability. Ferroelectric materials have been demonstrated to have memcapacitive properties. Yet, an ideal MFM structure has symmetric C-V characteristics (Figure. 3), leading to destructive readout in FeRAM, where this type of device is used. Non-intrinsic memcapacitive behaviour arises with interface or stack engineering, out of which different types of memcapacitive devices have emerged, namely, the MFM, MFS and FeFETs, as well as charge shielding devices. [67, 84-86]

As discussed in Section-2, ferroelectric materials show a net electrical polarization under applied electric field that remains after the electric field is removed. When an electric field is applied over the material, microscopic switching events start occurring. With increasing field magnitude, these nucleation sites start growing into domains, separated from other phases with domain walls. Once the polarization has fully switched, ideally no domain walls pertain in the material, and the polarization is in either of the two energetically favourable orientations. With the opposite direction of applied electric field, the polarization can be similarly switched to the opposite position. Charged defects and other nonidealities at the ferroelectric-metal interface can cause the domains to pin to one orientation, prohibiting domain rotation at those points.

The operation principle of MFM memcapacitors is based upon these purposefully included defect sites at only one of the metal-ferroelectric interfaces, as is visualized in Figure. 5. The charged defect sites forbid the dipole switching from that point, leading into coexistence of both polarized states in the material. Subsequently, some domain walls remain in only one of the saturated polarization states. This difference in the density of domain walls causes two different stable capacitance states, the LCS and HCS, due to the capacitance of the domain walls. The resulting asymmetric C-V curves enable the state of the capacitive device to be read with the benefits of memcapacitive devices.

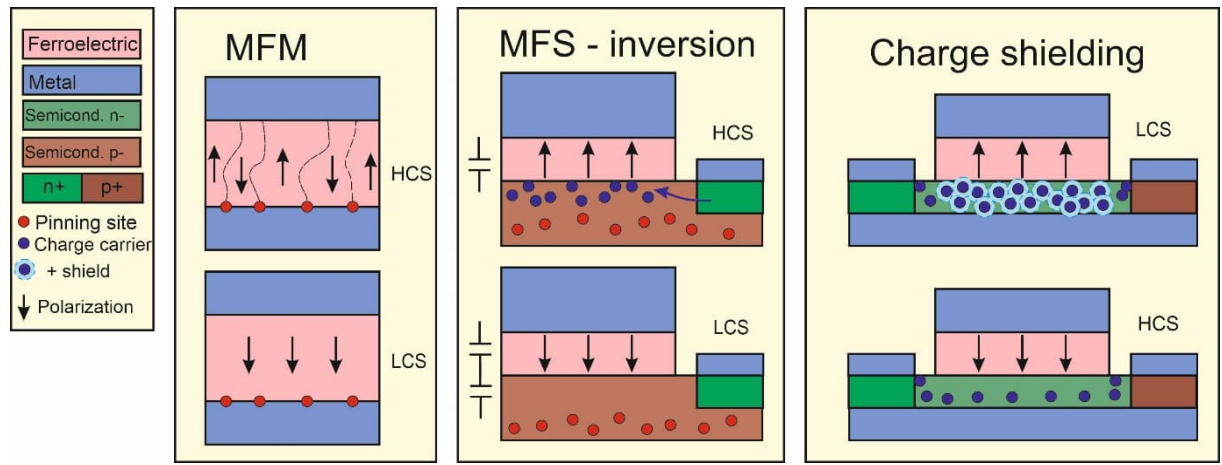

**Figure 5.** Device structure and operation principle of the three described ferroelectric memcapacitive devices.

Dependence of domain switching dynamics on the applied voltage pulse magnitude and temperature can be modelled utilizing physics-based models such as Jiles – Atherton model [87-88] that provides insight into the physical mechanism behind different kind of domain switching, asymmetric hysteresis under different pulsing conditions etc. providing insights on the ways to engineer devices for larger CMW. This can be seen from Figure. 6, where two different sets of dynamics parameters are compared. To achieve a high CMW at low operating voltages, it is desirable for the C-V response to begin increasing near zero voltage, rather than exhibiting a sharp peak at higher voltages. Simulation results indicate that reduced inter-domain coupling facilitates a more gradual and readable capacitance change, enhancing device sensitivity. Conversely, increased pinning loss also contributes to improved readability by promoting restricted domain switching, which initiates the C-V curve at lower voltages. Taken together, these findings suggest that partially crystallized, BEOL compatible annealing conditions may be more advantageous than fully crystalline states for the development of memcapacitive devices based on MFM structures. In fact, complete crystallinity might even hinder optimal CMW performance, whereas partial crystallization appears to strike a favourable balance between domain mobility and pinning, enabling enhanced capacitance modulation at low voltages.

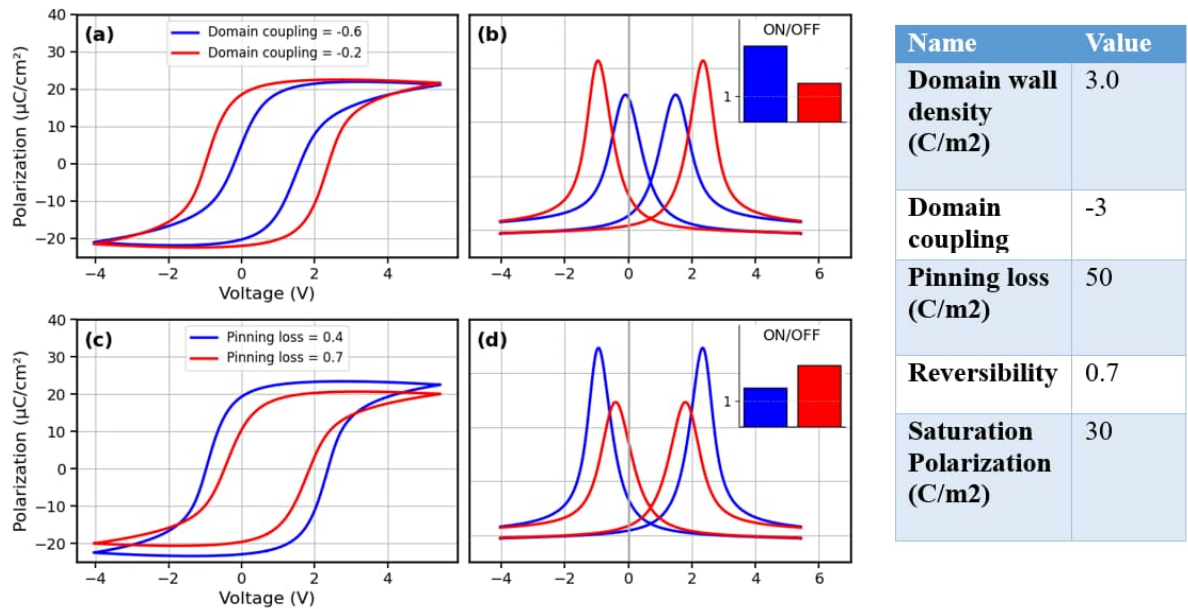

**Figure 6.** Simulated impact of domain dynamics parameters on the CWM. Here, an electrode work function difference of 0.7 eV was assumed, corresponding to asymmetric TiN/HZO/Au electrode structure.

The capacitance ratio between the HCS and LCS is limited to the noticeable, yet subtle difference in capacitance due to ferroelectric domain walls, and the capacitance ratios reached have been under 1.5, with a write voltage around 3 V. The small ON/OFF ratio causes small sensing margins, limits possible multi-level memories, and degrades the retention and endurance of the devices. [84] Due to these limitations, one possible future direction is to design material stacks with inherent switching asymmetry. Recent results have shown promising results for the polarization and endurance of these purposefully asymmetric devices. [89] Another focus development is a different operation mechanism with significantly enhanced capacitance ratio by introducing a semiconductor electrode onto the stack.

Next to a semiconductor interface, the charge imbalance of the polarization causes semiconductor charge carriers to either accumulate or deplete, depending on the direction of the polarization. When the semiconductor is in accumulation, only the ferroelectric layer contributes to the total capacitance. When a depletion region forms, it acts as a second capacitive element in series (as shown in Figure. 7), lowering the total capacitance of the stack, leading into the LCS of the MFS device. By switching the polarization, the device can switch between HCS and LCS, and the difference between these states is significantly increased compared to the MFM structure. This type of MFS device can be referred to as accumulation type MFS device, and the capacitance ratios obtained have been over 100 for wire voltages around 5 V, but additional optical stimulus is required. [90]

In the accumulation type MFS device, the number and generation rate of minority charge carriers in the semiconductor is insufficient without additional illumination to erase the wide depletion region formed during writing the LCS with a reasonable voltage amplitude. This erase issue has been addressed with a heavily doped minority carrier reservoir close to the ferroelectric-semiconductor interface, referred to as inversion type MFS memcapacitors [84], which is portrayed in Figure 5. The abundance of minority carriers assists in overcoming the very high erase voltage through inversion at the interface in the HCS. This structure also leads into a more symmetric switching between the conductance states at a lower voltage, and capacitance ratios of over 200 have been reported [84].

However, the write voltage of an MFS memcapacitor will always be higher than an MFM capacitor, due to the voltage drop in the depleted area of the semiconductor.

Utilizing the same operating principles as the MFS devices, FeFETs can also be used in a way to access the memcapacitive behaviour. By modulating the threshold voltage of the transistor, the capacitance state of the gate stack gets shifted, and capacitance ratios of over 20 have been achieved [45]. Some benefits of using FeFETs include the more mature knowledge of fabrication and optimization, as well as possibility of resistive reading. FeFETs have three terminals instead of two, making the design with them more complex compared to the two-terminal purely memcapacitive devices.

While the MFM memcapacitor is based upon the change in dielectric permittivity due to ferroelectric domain walls, and the MFS capacitor is based upon the charge screening in the semiconductor layer causing a depletion region to modulate the capacitance, a third way to control the capacitance is based on charge shielding. It requires a stack of two electrodes (gate and read-out), separated by a memory layer, in this case a ferroelectric layer, in addition to a lightly doped semiconductor layer acting as a shielding layer next to the read-out electrode. This stack is depicted in Figure 5. The shielding layer is, in addition, attached to a reservoir of electrons and holes, highly doped semiconductor regions connected to ground, to enable symmetric carrier injection [67]. The ferroelectric polarization is used to control whether the shielding layer semiconductor is in accumulation, depletion or inversion, similarly to the MFS memcapacitors.

Differing from MFS, where the change in capacitance is due to the depletion width, in charge shielding devices the carrier concentration affects the Debye screening. The higher the concentration of charge carriers in the shielding layer, the less electric field from the gate can reach the read-out terminal, resulting in a lower capacitance. Initially, both the polarization states correspond to a low capacitance, and the intermediary step in polarization switching where the shielding layer is near depletion leads into little to no shielding, and the screening length is long enough to reach the bottom read-out electrode. With informed device engineering, the gate electrode work function can be utilized to form a built-in bias field, causing one of the polarization directions to cause depletion in the shielding layer [91], leading into HCR. With the opposite polarization still attracting a high number of charge carriers, the shielding effect occurs, and the device remains in LCR.

The charge shielding type ferroelectric memcapacitor is a more complicated structure than the other types, though with the increased complexity, it allows for more fabrication parameters to be tuned as desired. The capacitive memory window of charge shielding capacitors depends on the shielding layer capacitance, which can be tuned with for example with the channel length and has been proposed to achieve levels of $10^3$ on/off with a small driving voltage of around 0.5 V [91], making charge shielding devices the most promising for neuromorphic capacitive systems where high on-off ratio is required.

*Improving Memory Retention*

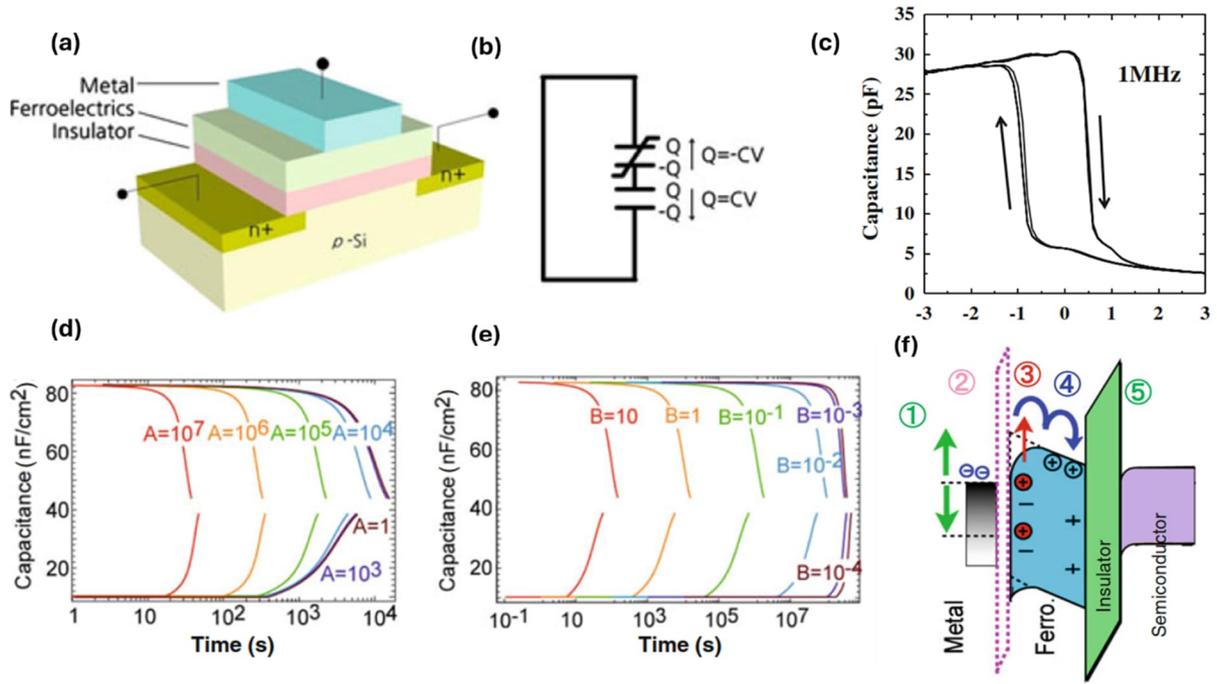

**Figure 7.** (a) An MFIS structure and its equivalent circuit showing (b) a linear and a non-linear capacitor at the gate stack controlling the (c) C-V characteristics of the device. Bottom panels show how decreasing current through the (d) insulator and (e) ferroelectric layers, improvement of memory retention can be attained. Capacitance retention characteristics for different proportionality factors A and B determine the currents across the semiconductor–insulator junction and through the metal-ferroelectric layers. (f) Various mechanisms for reducing leakage current through MFIS structure: 1) Decrease of metal work function. 2) Insertion of Insulator film. 3) Surface improvement of ferroelectric film. 4) Decrease of trap density. 5) High-k insulator film. (Figure reproduced with permission from [92])

To enhance the memory retention time, currents across the MFIS junction need to be reduced, which in ultra-thin ferroelectric films arises from several conduction mechanisms. Approaches for improvements are improving the metal-ferroelectric Shottky barrier, the ferroelectric layer itself, and the insulator-semiconductor junction. As a general rule, leakage currents across the Schottky junctions can be reduced by increasing the barrier height, as shown in Figure 7 (f). For instance, introduction of an electrode metal that increases the effective barrier height by almost 0.12 eV, can reduce the leakage current by a factor of $10^{-2}$, and thereby enhancing the retention time by a factor of $10^4$. However, increasing the barrier height by using the highest work function metal electrode is not a complete solution, since the Fermi level of a ferroelectric thin film tends to get pinned regardless of the metal's work function. Therefore, for improving the retention characteristics, the current across the Schottky junction can be suppressed by inserting an ultrathin insulator layer between the metal and ferroelectric layers. The inserted insulating layer should be as thin as possible, to minimize the depolarization field in the ferroelectric layer as well as the applied gate voltage for programming the states. A current passing through the ferroelectric layer encounters many traps that are produced during deposition and annealing treatments during the device fabrication. Trapping and releasing of carriers during the field cycling alter the charge distribution inside the device while charge-assisted carrier transport induces a space-charge-limited current. Reducing the trap density is therefore required to suppress the leakage current.

Current through the insulating layer and ferroelectric layer and their impact on retention behaviour is shown in Figure 7(d) and (e). Insulator film conduction criterion, A, can control retention as shown in Figure 7(d) and can be influenced by the real interface conditions such as changes in the barrier height and inhomogeneity of the Schottky junction. For the current factor through ferroelectric, the retention time can be extended by decreasing B to below $10^{-3}$. Comparing, we see that the retention characteristics are improved more by reducing current through ferroelectric than by reducing that through insulator. Therefore, it is desired that the current through the ferroelectric should be minimized as far as possible. Figure 7(f) shows various mechanisms for reducing leakage current through MFIS structure such as decrease of metal work function, insertion of insulator film, surface improvement of ferroelectric film, decrease of trap density and high-k insulator film that can have significant improvement on the retention behaviour.

Overall, ferroelectric capacitive synapses offer several key advantages for neuromorphic compute-in-memory systems. MFS inversion and charge shielding type devices provide a high on/off capacitance ratio, enabling clear distinction between stored states for accurate readout. Since memcapacitive devices use small-signal AC sensing, the readout is non-destructive, avoiding issues like read disturbance seen in resistive memories. Their high-impedance nature eliminates sneak-path currents, allowing selector-less operation and simplifying array design. Additionally, they consume virtually no static power, making them highly energy efficient. The devices are also compatible with back-end-of-line (BEOL) integration, supporting dense 3D stacking above CMOS circuitry for compact, high-performance hardware architectures [16]. Nevertheless, relatively large footprint and integration complexity are trade-offs to consider for large-scale neuromorphic CIM arrays. Future developments regarding ferroelectric memcapacitors include increasing the capacitive memory window ensuring multi-level operation.

## 5. Bridging Device Physics and System Metrics

While structural and material innovations in capacitive synapses, such as MFM, MFS, FeFET and charge-shielding devices, enable enhancements in CMW and other key device parameters, their impact on system-level performance is governed by intricate trade-offs. Device-level behaviours directly influence critical metrics such as noise characteristics, D2D variation, charge transfer fidelity, and the effective number of bits (ENOB) achievable in capacitive CIM arrays.

*Impact of device-Level Phenomena on system performance*

FeCAP devices inherently exhibit nonidealities arising from inherent stochasticity in domain dynamics as well as interfacial and bulk trap states. These effects manifest as switching nonlinearities, state-dependent retention (e.g., imprint or offset in the P–E loop) [93] and contribute to random telegraph noise (RTN) and gradual drift in the effective capacitance. Multi-domain switching, hysteresis, and charge trapping cumulatively lead to significant device-level variability, which must be thoroughly understood before system-level deployment.

These intrinsic phenomena collectively influence essential synaptic device properties, including linearity, asymmetry, on/off ratio, weight precision, retention, and endurance. For example, the linearity of analog MAC operations is compromised by ferroelectric hysteresis and domain saturation effects. Ideally, the accumulated charge should vary linearly with synaptic weight training, but saturation effects limit this, reducing the usable dynamic range and consequently the ENOB [69]. Similarly, interfacial trap-induced noise elevates the analog noise floor, degrading computational precision to ~2–4 bits unless compensated. Reliable neuromorphic systems thus require balanced optimization across all these parameters (Figure. 8).

Latency and throughput are also strongly affected by device dynamics. In FeCAP crossbars, read/write operations consist of two distinct phases: charging and sensing phase as discussed in section-

3. [70]. This inherently multi-phase operation, coupled with additional verify-and-program schemes to manage hysteresis, increases computational latency. Furthermore, stochastic domain switching and trap-release phenomena contribute to variability in output analog sums, compounding errors at scale and severely impacting the throughput.

While adopting an analog charge-collection approach in FeCAP crossbars can alleviate ADC (Analog-to-Digital Converter) precision requirements, non-idealities such as noise, charge leakage, non-linearity, and cycle-to-cycle or device-to-device variability must be carefully managed to avoid greater analog circuit complexity (as discussed later).

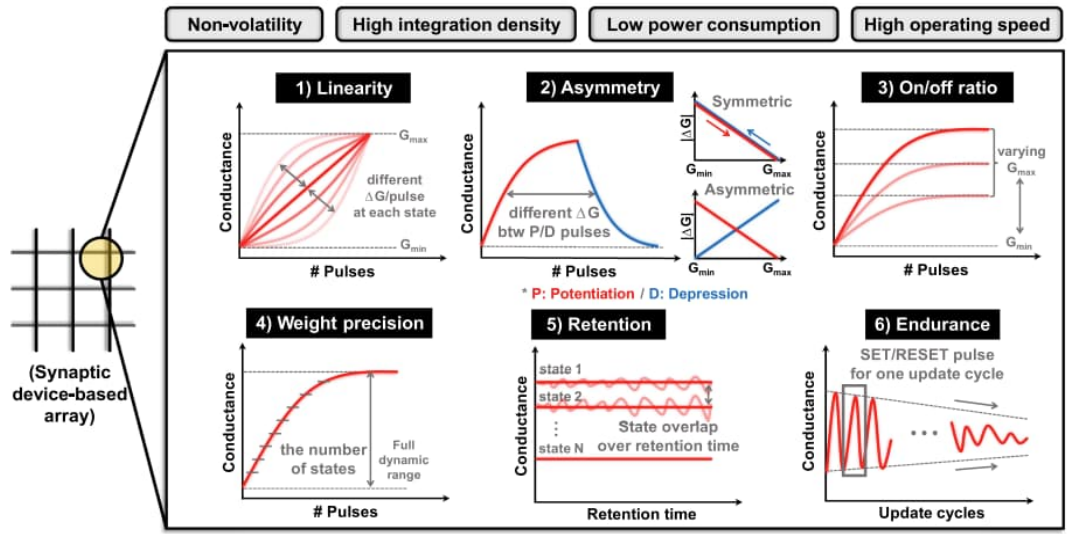

**Figure 8.** Key device-level performance metrics required for large-scale neuromorphic computing systems. The synaptic device-based array must demonstrate non-volatility, high integration density, low power consumption, and high operating speed. Critical characteristics include: (1) Linearity of conductance updates per pulse, (2) Asymmetry between potentiation and depression responses, (3) Sufficient on/off ratio for distinct state differentiation, (4) High weight precision for achieving a large number of programmable conductance states, (5) Retention stability to preserve synaptic weights over time, and (6) Endurance against repeated SET/RESET cycling during training and inference (Figure reproduced with permission from [94]).

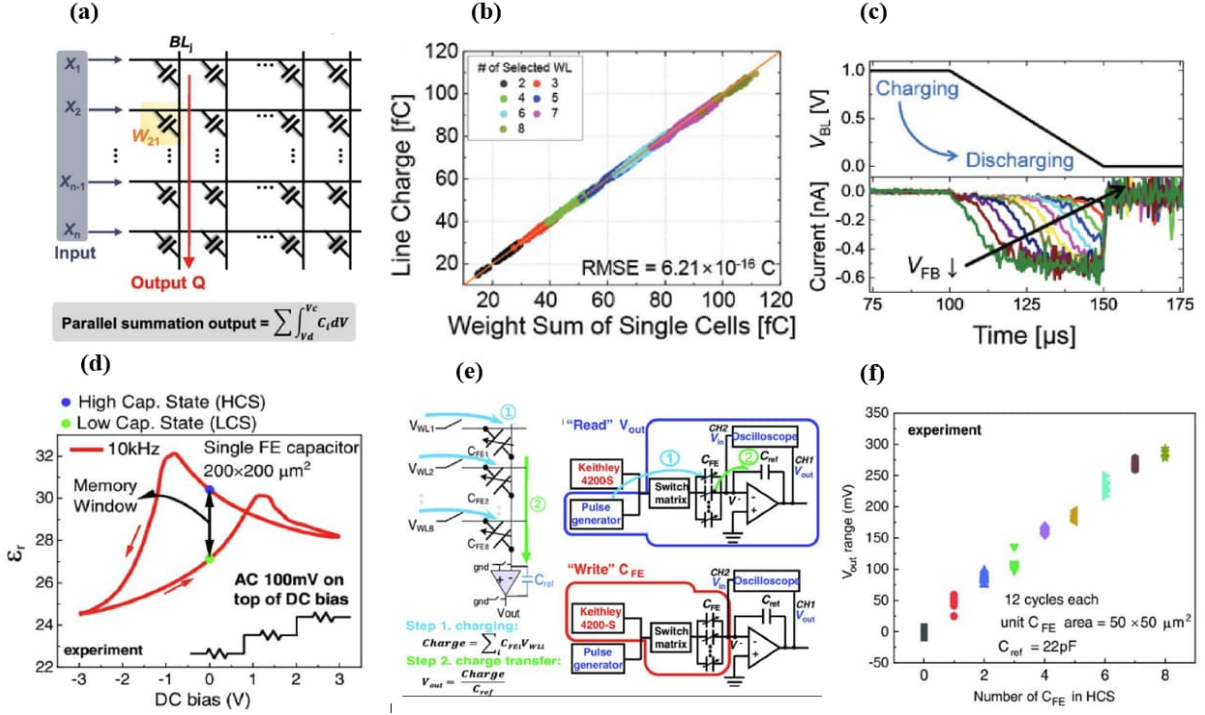

**Figure 9.** Capacitance-Based Devices and Array Structures for In-Memory Computing. **(a)** Array schematic enabling charge-based MAC. **(b)** High linearity of line charge vs. weighted sum. **(c)** Bitline charging/discharging dynamics. **(d)** FeCAP hysteresis showing memory window. **(e)** Peripheral circuit for read/write via charge transfer. **(f)** Scalable $V_{out}$ with increasing FeCAPs in HCS. (Figure reproduced with permission from [94]).

*Circuit-Level Mitigation Strategies*

The overall system of a small-scale capacitive CIM architecture can be seen in Fig. 9. It can be observed that the read/write circuitry (e) uses a reference capacitor and charge-transfer mechanism to sense weights accurately, enabling analog-to-digital conversion through voltage integration. The linear charge accumulation demonstrated in (b) confirm high computational fidelity, while the multilevel programmable states should support the fine-grained weight encoding. Charging dynamics in (c) and voltage scalability in (f) highlight the robustness of the peripheral readout path, making the design well-suited for high-resolution, energy-efficient neuromorphic computing. However, this kind of performance cannot be maintained with scaling up of the crossbar structure. Following techniques adopted in previous studies such as [69] made an attempt to mitigate the limitations offered by the system design.

Several circuit-level techniques have been proposed to mitigate these device-level limitations. A key approach is correlated double sampling (CDS), where the baseline signal is sampled prior to charge integration and then subtracted from the post-integration signal. This technique cancels out fixed offsets and low-frequency noise (e.g., trap-induced drifts, amplifier 1/f noise), thus enhancing the effective precision and linearity of analog readout [69].

Another strategy is the deployment of offset-cancelled or offset-tuned amplifiers. Here, programmable calibration or dummy reference FeCAPs are employed to nullify systematic offsets caused by device imprint or mismatch. This ensures that nominal zero-weight conditions correspond to zero output, maximizing dynamic range and boosting the ENOB of crossbar computations [69].

FeCAP crossbars also benefit from feedback-based sensing architectures. Unlike resistive crossbars that suffer from IR drop and non-uniform biasing, FeCAP arrays leverage charge-mode operation, avoiding such issues but still requiring feedback control to maintain consistent operating points [70]. A transimpedance amplifier (TIA) or integrator at the bitline can hold the node at virtual ground via negative feedback. During readout, charge is accumulated in the integrator, maintaining linear proportionality to the stored polarization. By integrating charge onto a feedback capacitor until a voltage threshold is reached, analog-to-digital conversion can be achieved with minimal sensitivity to noise, allowing high linearity and noise suppression. Additional enhancements include differential sensing—using paired FeCAPs to represent positive and negative weights, thereby cancelling common-mode noise—and autozeroing techniques to periodically reset amplifier offsets. Combined, these circuit-level methods significantly mitigate the effects of nonlinearity, offset drift, and random telegraph noise, pushing the effective system precision closer to the theoretical limits set by fundamental device variability [69, 70].

*Requirement of cross-layer Co-Optimization in Neuromorphic CIM*

Despite these advances, no single circuit-level solution fully compensates for all device nonidealities. Hence, a cross-layer co-optimization strategy is imperative for realizing robust FeCAP-based neuromorphic systems. At the device level, material engineering techniques, such as precise doping and annealing of $HfO_2$-based ferroelectrics, can be employed to achieve lower coercive fields, minimized hysteresis, and reproducible domain switching, thereby improving weight update linearity and retention [94].

Concurrently, circuit designers can focus on developing adaptive read/write protocols, such as program-and-verify algorithms, to linearize synaptic updates. Peripheral circuits employing CDS, feedback integration, and offset cancellation are carefully optimized to extract maximum precision. At the architectural level, computational strategies like bit-slicing or time-multiplexed input encoding reduce the burden on instantaneous analog precision by distributing high-bit significance computations across multiple low-precision cycles [70].

Hardware-aware training algorithms might further complement this stack, allowing the network to adapt to nonideal device behaviors, including hysteresis and drift. Error-correcting codes or redundancy schemes (e.g., averaging multiple FeCAP cells per weight) are also deployed to suppress residual variability. Cross-layer improvements synergistically enhance system performance: cleaner device physics simplifies circuit design, and improved circuits allow larger, denser crossbar arrays without sacrificing accuracy.

Recent experimental studies demonstrate that such co-optimized FeCAP crossbar systems can achieve significant gains in end-to-end inference accuracy. For instance, workloads like CIFAR-10 have reported accuracy improvements from near random baselines to competitive performance levels (~80–90%) after applying combined device, circuit, and algorithmic optimizations [69]. In conclusion, bridging device physics with circuit and system design is essential for unlocking the full potential of FeCAP-based crossbar arrays in energy-efficient, high-accuracy neuromorphic computing platforms. A quick comparison of the various candidates of the capacitive memories can be viewed in Table. 2.

**Table 2:** Comparison of capacitive synaptic Device types

| Device Type | Reference | Physical Mechanism | Capacitance Ratio (CH/CL) | Write Voltage (V) | Area ($F^2$) | ENOB (bits) | Key Advantages | Key Limitations |
|---|---|---|---|---|---|---|---|---|
| MFM (Metal-Ferroelectric-Metal) | [84] | Domain wall displacement | ~1.15 | ±3 | ~4 | 2–3 | Simple structure, robust reliability | Low capacitance ratio, limited dynamic range |
| MFS (Metal-Ferroelectric-Semiconductor) | [84, 85]] | Depletion modulation of ferroelectric polarization | 24–208 | ±2.5 to ±5.5 | ~4 | 3–5 | CMOS compatibility, moderate linearity | Interface sensitivity, doping-dependent behavior |
| FeFET (Ferroelectric FET) | [86] | Polarization-controlled channel conduction | ~25 | ±3.5 | 6–10 | 4–6 | In-built gain, digital and analog compatibility | Subthreshold leakage, hysteresis variability |
| Charge Shielding Capacitor | [67] | Ferroelectric modulation + electrostatic shielding | 60–90 | ±3 | ~16 | 5–6 | Highest capacitance ratio, sneak-path suppression | Sensitive to interfacial traps, doping gradients |

**Conclusion and outlook**

This work has provided a holistic perspective on emerging synaptic memory technologies for neuromorphic CIM platforms, beginning with a comparative analysis of resistive memories, such as RRAM, PCM, MRAM, and FRAM, and culminating in a focused exploration of capacitive memory devices. While resistive memories offer mature device integration and multilevel storage, they face fundamental challenges such as sneak-path currents, high static power, read disturbance, and endurance degradation under repeated switching. These limitations motivate the exploration of alternative paradigms such as capacitive synapses, particularly those leveraging ferroelectric materials. We presented detailed structural and operational insights into various capacitive synapse architectures, including MFM, MFS, FeFET, and charge-shielding devices. These devices exploit electric-field-driven polarization dynamics and charge modulation mechanisms to encode synaptic weights. Their operation enables low-power, selectorless array integration and inherently avoids the issues of resistive heating and filamentary randomness encountered in RRAM-based systems. Moreover, the detailed discussion on charge integration schemes, read/write phases, and analog accumulation mechanisms highlights their suitability for parallel analog MAC operations.

Nevertheless, capacitive memories are not free from device-level nonidealities. Phenomena such as ferroelectric hysteresis, domain saturation, interfacial traps, and charge retention drifts introduce variability that degrades precision and limits effective resolution (ENOB). We touched upon how these effects influence critical system parameters, such as linearity, on/off ratio, weight update symmetry, and read latency, and how circuit-level mitigation techniques (e.g., CDS, offset cancellation, differential sensing) play a central role in preserving computational fidelity. Ultimately, the key to unlocking the potential of capacitive CIM systems lies in cross-layer co-optimization. This includes material engineering for stable domain dynamics, circuit innovations for precise analog signal processing, architectural adaptations for redundancy and bit-slicing, and hardware-aware algorithms capable of learning under imperfect conditions. Experimental results show that such synergy can elevate inference accuracy in real-world tasks to competitive levels while maintaining energy efficiency and scalability.

In conclusion, capacitive synapses, with their low static power, minimal read disturbance, and CMOS BEOL compatibility, offer a fundamentally robust alternative to resistive memories. As device fabrication matures and system-level co-design methodologies evolve, ferroelectric capacitive memories can stand out as highly promising candidates for next-generation neuromorphic hardware, enabling compact, energy-efficient, and high-accuracy in-memory computing architectures.

**References**

[1] M. Capra, R. Peloso, G. Masera, M. Ruo Roch, and M. Martina, "Edge computing: A survey on the hardware requirements in the internet of things world," *Future Internet* 11, no. 4 (2019): 100.

[2] A. Mehonic, D. Ielmini, K. Roy, O. Mutlu, S. Kvatinsky, T. Serrano-Gotarredona, et al., "Roadmap to neuromorphic computing with emerging technologies," *APL Materials* 12, no. 10 (2024).

[3] S. Yu, "Neuro-inspired computing with emerging nonvolatile memorys," *Proceedings of the IEEE* 106, no. 2 (2018): 260–285.

[4] W. Haensch, A. Raghunathan, K. Roy, B. Chakrabarti, C. M. Phatak, C. Wang, and S. Guha, "Compute in-memory with non-volatile elements for neural networks: A review from a co-design perspective," *Advanced Materials* 35, no. 37 (2023): 2204944.

[5] Y. Chen, "ReRAM: History, status, and future," *IEEE Transactions on Electron Devices* 67, no. 4 (2020): 1420–1433.

[6] D. Strukov, G. Snider, D. Stewart, and R. S. Williams, "The missing memristor found," *Nature* 453 (2008): 80–83. https://doi.org/10.1038/nature06932

[7] Z. Biolek, D. Biolek, and V. Biolkova, "SPICE model of memristor with nonlinear dopant drift," *Radioengineering* 18, no. 2 (2009).

[8] O. Kavehei, A. Iqbal, Y. S. Kim, K. Eshraghian, S. F. Al-Sarawi, and D. Abbott, "The fourth element: characteristics, modelling and electromagnetic theory of the memristor," *Proceedings of the Royal Society A: Mathematical, Physical and Engineering Sciences* 466, no. 2120 (2010): 2175–2202.

[9] P. Sheridan, K. H. Kim, S. Gaba, T. Chang, L. Chen, and W. Lu, "Device and SPICE modeling of RRAM devices," *Nanoscale* 3, no. 9 (2011): 3833–3840.

[10] Z. Shen, C. Zhao, Y. Qi, W. Xu, Y. Liu, I. Z. Mitrovic, ... and C. Zhao, "Advances of RRAM devices: Resistive switching mechanisms, materials and bionic synaptic application," *Nanomaterials* 10, no. 8 (2020): 1437.

[11] T. Kim and S. Lee, "Evolution of phase-change memory for the storage-class memory and beyond," *IEEE Transactions on Electron Devices* 67, no. 4 (2020): 1394–1402.

[12] S. Fukami and H. Ohno, "Perspective: Spintronic synapse for artificial neural network," *APL Materials* 4 (2016): 032503.

[13] J. K. De Brosse, L. Liu, and D. Worledge, "Spin Hall effect assisted spin transfer torque magnetic random access memory," U.S. Patent No. 8,896,041 (25 Nov. 2014).

[14] E. M. Philofsky, "FRAM-the ultimate memory," in *Proceedings of Nonvolatile Memory Technology Conference*, Albuquerque, NM, USA, 1996, 99–104. https://doi.org/10.1109/NVMT.1996.534679

[15] R. Athle and M. Borg, "Ferroelectric tunnel junction memristors for in-memory computing accelerators," *Advanced Intelligent Systems* 6, no. 3 (2024): 2300554.

[16] S. Majumdar, "Can ferroelectric tunnel junction be a game changer as eNVM and in neuromorphic hardware?", *APL Machine Learning* 3, no. 2 (2025).

[17] S. Majumdar, H. Tan, I. Pande, and S. Van Dijken, "Crossover from synaptic to neuronal functionalities through carrier concentration control in Nb-doped SrTiO3-based organic ferroelectric tunnel junctions," *APL Materials* 7, no. 9 (2019).

[18] I. J. Kim and J. S. Lee, "Ferroelectric transistors for memory and neuromorphic device applications," *Advanced Materials* 35, no. 22 (2023): 2206864.

[19] S. Majumdar and I. Zeimpekis, "Back-end and flexible substrate compatible analog ferroelectric field-effect transistors for accurate online training in deep neural network accelerators," *Advanced Intelligent Systems* 5, no. 11 (2023): 2300391.

[20] E. Paasio, R. Ranta, and S. Majumdar, "A physics-based compact model for ferroelectric capacitors operating down to deep cryogenic temperatures for applications in analog memory and neuromorphic architectures," *Advanced Electronic Materials* 11, no. 9 (2025): 2400840.

[21] R. Pelke, J. Cubero-Cascante, N. Bosbach, N. Degener, F. Idrizi, L. M. Reimann, ... and R. Leupers, "Optimizing binary and ternary neural network inference on RRAM crossbars using CIM-Explorer," *arXiv preprint* arXiv:2505.14303 (2025).

[22] H. Jeong, S. Kim, K. Park, J. Jung, and K. J. Lee, "A ternary neural network computing-in-memory processor with 16T1C bitcell architecture," *IEEE Transactions on Circuits and Systems II: Express Briefs* 70, no. 5 (2023): 1739–1743.

[23] S. Majumdar, "An efficient deep neural network accelerator using controlled ferroelectric domain dynamics," *Neuromorphic Computing and Engineering* 2, no. 4 (2022): 041001.

[24] K. Bhardwaj and M. Srivastava, "Floating memristor and inverse memristor emulation configurations with electronic/resistance controllability," *IET Circuits, Devices & Systems* 14, no. 7 (2020): 1065–1076.

[25] D. Biolek, Z. Kohl, J. Vavra, V. Biolková, K. Bhardwaj, and M. Srivastava, "Mutual transformation of flux-controlled and charge-controlled memristors," *IEEE Access* 10 (2022): 68307–68318.

[26] D. Sacchetto, M. A. Doucey, G. De Micheli, Y. Leblebici, and S. Carrara, "New insight on bio-sensing by nano-fabricated memristors," *BioNanoScience* 1 (2011): 1–3.

[27] K. Bhardwaj, D. Semenov, R. Sotner, J. Chen, S. Carrara, and M. Srivastava, "Emulating multimemristive behavior of silicon nanowire-based biosensors by using CMOS-based implementations," *IEEE Sensors Journal* 24, no. 6 (2024): 8036–8044.

[28] W. Wan, R. Kubendran, C. Schaefer, S. B. Eryilmaz, W. Zhang, D. Wu, ... and G. Cauwenberghs, "A compute-in-memory chip based on resistive random-access memory," *Nature* 608, no. 7923 (2022): 504–512.

[29] P. Yao, H. Wu, B. Gao, J. Tang, Q. Zhang, W. Zhang, ... and H. Qian, "Fully hardware-implemented memristor convolutional neural network," *Nature* 577, no. 7792 (2020): 641–646.

[30] H. Kim, M. R. Mahmoodi, H. Nili, ... et al., "4K-memristor analog-grade passive crossbar circuit," *Nature Communications* 12 (2021): 5198. https://doi.org/10.1038/s41467-021-25455-0

[31] M. Rao, H. Tang, J. Wu, ... et al., "Thousands of conductance levels in memristors integrated on CMOS," *Nature* 615 (2023): 823–829.

[32] H. Aziza, "Image classification in memristor-based neural networks: A comparative study of software and hardware models using RRAM crossbars," *Electronics* 14, no. 6 (2025): 1125.

[33] M. D. Gomony, B. Ahn, R. Luiken, Y. Biyani, A. Gebregiorgis, A. Laborieux, ... and H. Corporaal, "Achieving PetaOps/W Edge-AI processing," in *Proceedings of the 61st ACM/IEEE Design Automation Conference*, June 2024, pp. 1–4.

[34] S. Liu, S. Wei, P. Yao, D. Wu, L. Jie, S. Pan, ... and H. Wu, "A 28 nm 576K RRAM-based computing-in-memory macro featuring hybrid programming with area efficiency of 2.82 TOPS/mm²," *Journal of Semiconductors* 46 (2025): 1–8.

[35] M. Ezzadeen, A. Majumdar, O. Valorge, N. Castellani, V. Gherman, G. Regis, ... and J. M. Portal, "Implementation of binarized neural networks immune to device variation and voltage drop employing resistive random access memory bridges and capacitive neurons," *Communications Engineering* 3, no. 1 (2024): 80.

[36] J. B. Roldán, E. Miranda, D. Maldonado, A. N. Mikhaylov, N. V. Agudov, A. A. Dubkov, ... and L. O. Chua, "Variability in resistive memories," *Advanced Intelligent Systems* 5, no. 6 (2023): 2200338.

[37] Y. Cassuto, S. Kvatinsky, and E. Yaakobi, "Sneak-path constraints in memristor crossbar arrays," in *2013 IEEE International Symposium on Information Theory*, Istanbul, Turkey, 2013, pp. 156–160. https://doi.org/10.1109/ISIT.2013.6620207

[38] X. X. Xu, Q. Luo, T. C. Gong, H. B. Lv, Q. Liu, and M. Liu, "Resistive switching memory for high density storage and computing," *Chinese Physics B* 30, no. 5 (2021): 058702.

[39] J. Zhou, K.-H. Kim, and W. Lu, "Crossbar RRAM arrays: Selector device requirements during read operation," *IEEE Transactions on Electron Devices* 61, no. 5 (2014): 1369–1376.

[40] S. Mukherjee, J. Bizindavyi, S. Clima, M. I. Popovici, X. Piao, K. Katcko, ... and J. Van Houdt, "Capacitive memory window with non-destructive read in ferroelectric capacitors," *IEEE Electron Device Letters* 44, no. 7 (2023): 1092–1095.

[41] M. Di Ventra, Y. V. Pershin, and L. O. Chua, "Circuit elements with memory: Memristors, memcapacitors, and meminductors," *Proceedings of the IEEE* 97, no. 10 (2009): 1717–1724.

[42] K. Ni, X. Yin, A. F. Laguna, ... et al., "Ferroelectric ternary content-addressable memory for one-shot learning," *Nature Electronics* 3 (2020): 130. https://doi.org/10.1038/s41928-020-0374-3

[43] D. Lehninger, M. Lederer, T. Ali, T. Kämpfe, K. Mertens, and K. Seidel, "Enabling ferroelectric memories in BEoL – towards advanced neuromorphic computing architectures," in *2021 IEEE International Interconnect Technology Conference (IITC)*, Kyoto, Japan, 2021.

[44] S. Majumdar, "Back-end CMOS compatible and flexible ferroelectric memories for neuromorphic computing and adaptive sensing," *Advanced Intelligent Systems* 4, no. 4 (2022): 2100175.

[45] S. Yu, Y.-C. Luo, T.-H. Kim, and O. Phadke, "Nonvolatile capacitive synapse device candidates for charge domain compute-in-memory," *IEEE Electron Devices Magazine* 9, no. 3 (2023): 23–32.

[46] O. Phadke, T. H. Kim, Y. C. Luo, and S. Yu, "Ferroelectric nonvolatile capacitive synapse for charge domain compute-in-memory," *ECS Transactions* 113, no. 14 (2024): 3.

[47] R. Alcala, M. Materano, P. D. Lomenzo, L. Grenouillet, T. Francois, J. Coignus, N. Vaxelaire, ... et al., "BEOL integrated ferroelectric $HfO_2$-based capacitors for FeRAM: Extrapolation of reliability performance to use conditions," *IEEE Journal of the Electron Devices Society* 10 (2022): 907–912.

[48] M. A. Lastras-Montaño and K. T. Cheng, "Resistive random-access memory based on ratioed memristors," *Nature Electronics* 1 (2018): 466–472. https://doi.org/10.1038/s41928-018-0115-z

[49] K. Jeon, J. Kim, J. J. Ryu, ... et al., "Self-rectifying resistive memory in passive crossbar arrays," *Nature Communications* 12 (2021): 2968. https://doi.org/10.1038/s41467-021-23180-2

[50] M. Le Gallo and A. Sebastian, "An overview of phase-change memory device physics," *Journal of Physics D: Applied Physics* 53, no. 21 (2020): 213002.

[51] G. S. Syed, M. Le Gallo, and A. Sebastian, "Phase-change memory for in-memory computing," *Chemical Reviews* (2025).

[52] S. Ambrogio, P. Narayanan, A. Okazaki, ... et al., "An analog-AI chip for energy-efficient speech recognition and transcription," *Nature* 620 (2023): 768–775. https://doi.org/10.1038/s41586-023-06337-5

[53] M. M. Frank, N. Li, M. J. Rasch, S. Jain, C. T. Chen, R. Muralidhar, ... and G. W. Burr, "Impact of phase-change memory drift on energy efficiency and accuracy of analog compute-in-memory deep learning inference," in *2023 IEEE International Reliability Physics Symposium (IRPS)*, March 2023, pp. 1–10.

[54] B. N. Engel, J. Akerman, B. Butcher, R. Dave, M. DeHerrera, M. Durlam, ... et al., "The science and technology of magnetoresistive tunneling memory," *IEEE Transactions on Nanotechnology* 1, no. 1 (2002): 32–38. https://doi.org/10.1109/TNANO.2002.1005424

[55] G. Prenat, K. Jabeur, P. Vanhauwaert, G. Di Pendina, F. Oboril, R. Bishnoi, ... and G. Gaudin, "Ultra-fast and high-reliability SOT-MRAM: From cache replacement to normally-off computing," *IEEE Transactions on Multi-Scale Computing Systems* 2, no. 1 (2015): 49–60.

[56] R. Saha, Y. P. Pundir, and P. K. Pal, "Comparative analysis of STT and SOT based MRAMs for last level caches," *Journal of Magnetism and Magnetic Materials* 551 (2022): 169161.

[57] S. Jung, H. Lee, S. Myung, ... et al., "A crossbar array of magnetoresistive memory devices for in-memory computing," *Nature* 601 (2022): 211–216.

[58] S. Majumdar, "Ultrafast switching and linear conductance modulation in ferroelectric tunnel junctions via P(VDF–TrFE) morphology control," *Nanoscale* 13, no. 25 (2021): 11270–11278.

[59] X. Li, P. Srivari, E. Paasio, and S. Majumdar, "Understanding fatigue and recovery mechanisms in $Hf_{0.5}Zr_{0.5}O_2$ capacitors for designing high endurance ferroelectric memory and neuromorphic hardware," *Nanoscale* 17, no. 10 (2025): 6058–6071.

[60] P. Srivari, E. Paasio, X. Li, and S. Majumdar, "Towards improved polarization uniformity in ferroelectric $Hf_{0.5}Zr_{0.5}O_2$ devices within back end of line thermal budget for memory and neuromorphic applications," *arXiv preprint* arXiv:2412.11288 (2024).

[61] J. Hur, C. Park, G. Choe, P. V. Ravindran, A. I. Khan, and S. Yu, "Characterizing $HfO_2$-based ferroelectric tunnel junction in cryogenic temperature," *IEEE Transactions on Electron Devices* 69, no. 10 (2022): 5948–5951.

[62] H. Bohuslavskyi, K. Grigoras, M. Ribeiro, M. Prunnila, and S. Majumdar, "Ferroelectric $Hf_{0.5}Zr_{0.5}O_2$ for analog memory and in-memory computing applications down to deep cryogenic temperatures," *Advanced Electronic Materials* 10, no. 7 (2024): 2300879.

[63] A. F. Laguna, X. Yin, D. Reis, M. Niemier, and X. S. Hu, "Ferroelectric FET-based in-memory computing for few-shot learning," in *Proceedings of the 2019 Great Lakes Symposium on VLSI* (2019): 373–378.

[64] S. Dutta, A. Khanna, H. Ye, M. M. Sharifi, A. Kazemi, M. San Jose, ... and S. Datta, "Lifelong learning with monolithic 3D ferroelectric ternary content-addressable memory," in *2021 IEEE International Electron Devices Meeting (IEDM)* (2021): 1–4.

[65] K. Seidel, D. Lehninger, A. Sünbül, R. Hoffmann, R. Revello, N. Yadav, ... and M. Lederer, "Integration of ferroelectric devices for advanced in-memory computing concepts," *Japanese Journal of Applied Physics* 63, no. 5 (2024): 050802.

[66] Z. Zhou, Y. Zhou, S. Kang, X. Wang, Y. Yu, and H. Qian, "A metal–insulator–semiconductor non-volatile programmable capacitor based on a $HfAlO_x$ ferroelectric film," *IEEE Electron Device Letters* 41, no. 12 (2020): 1837–1840.

[67] K. U. Demasius, A. Kirschen, and S. Parkin, "Energy-efficient memcapacitor devices for neuromorphic computing," *Nature Electronics* 4, no. 10 (2021): 748–756.

[68] Y. C. Luo, A. Lu, J. Hur, S. Li, and S. Yu, "Design and optimization of non-volatile capacitive crossbar array for in-memory computing," *IEEE Transactions on Circuits and Systems II: Express Briefs* 69, no. 3 (2021): 784–788.

[69] E. Yu, G. K. K., U. Saxena, and K. Roy, "Ferroelectric capacitors and field-effect transistors as in-memory computing elements for machine learning workloads," *Scientific Reports* 14, no. 1 (2024): 9426.

[70] J. Hur, Y. C. Luo, A. Lu, T. H. Wang, S. Li, A. I. Khan, and S. Yu, "Nonvolatile capacitive crossbar array for in-memory computing," *Advanced Intelligent Systems* 4, no. 8 (2022): 2100258.

[71] X. Sun, W. S. Khwa, Y. S. Chen, C. H. Lee, H. Y. Lee, S. M. Yu, ... and K. Akarvardar, "PCM-based analog compute-in-memory: Impact of device non-idealities on inference accuracy," *IEEE Transactions on Electron Devices* 68, no. 11 (2021): 5585–5591.

[72] H. Cai, Y. Guo, B. Liu, M. Zhou, J. Chen, X. Liu, and J. Yang, "Proposal of analog in-memory computing with magnified tunnel magnetoresistance ratio and universal STT-MRAM cell," *IEEE Transactions on Circuits and Systems I: Regular Papers* 69, no. 4 (2022): 1519–1531.

[73] C. Chen, L. Goux, A. Fantini, A. Redolfi, S. Clima, R. Degraeve, et al., "Understanding the impact of programming pulses and electrode materials on the endurance properties of scaled $Ta_2O_5$ RRAM cells," in *2014 IEEE International Electron Devices Meeting* (2014).

[74] R. Waser, "Resistive non-volatile memory devices," *Microelectronic Engineering* 86, no. 7–9 (2009): 1925–1928.

[75] D. Ielmini, "Brain-inspired computing with resistive switching memory (RRAM): Devices, synapses and neural networks," *Microelectronic Engineering* 190 (2018): 44–53.

[76] Z. Zhou, L. Jiao, J. Zhou, Z. Zheng, Y. Chen, K. Han, ... and X. Gong, "Inversion-type ferroelectric capacitive memory and its 1-kbit crossbar array," *IEEE Transactions on Electron Devices* 70, no. 4 (2023): 1641–1647.

[77] T. Soliman, S. Chatterjee, N. Laleni, et al., "First demonstration of in-memory computing crossbar using multi-level cell FeFET," *Nature Communications* 14 (2023): 6348. https://doi.org/10.1038/s41467-023-42110-y

[78] M. Hellenbrand, I. Teck, and J. L. MacManus-Driscoll, "Progress of emerging non-volatile memory technologies in industry," *MRS Communications* (2024): 1–14.

[79] H. Lin and Y. Jiang, "Optimization of the 3D multi-level SOT-MRAMs," *AIP Advances* 14, no. 2 (2024).

[80] W. S. Khwa, Y. L. Lu, S. Q. Zhang, X. Sun, S. S. Sarwar, Z. Li, ... and M. F. Chang, "MRAM design–technology–system co-optimization for artificial intelligence edge devices," in *2024 IEEE International Electron Devices Meeting (IEDM)* (2024): 1–4.

[81] G. W. Burr, M. J. Breitwisch, M. Franceschini, D. Garetto, K. Gopalakrishnan, B. Jackson, ... and R. S. Shenoy, "Phase change memory technology," *Journal of Vacuum Science & Technology B* 28, no. 2 (2010): 223–262.

[82] H. Y. Cheng, F. Carta, W. C. Chien, H. L. Lung, and M. J. BrightSky, "3D cross-point phase-change memory for storage-class memory," *Journal of Physics D: Applied Physics* 52, no. 47 (2019): 473002.

[83] D. J. Wouters, R. Waser, and M. Wuttig, "Phase-change and redox-based resistive switching memories," *Proceedings of the IEEE* 103, no. 8 (2015): 1274–1288.

[84] Z. Zhou, L. Jiao, Z. Zheng, et al., "Ferroelectric capacitive memories: devices, arrays, and applications," *Nano Convergence* 12, 3 (2025). https://doi.org/10.1186/s40580-024-00380-0

[85] J. Lee, K. Lee, D. Kim, J. Park, T. Kim, and H. Park, "BEOL-compatible non-volatile capacitive synapse with ALD W-doped $In_2O_3$ semiconductor layer," in *2024 IEEE International Electron Devices Meeting (IEDM)* (2024): 1–4.

[86] T.-H. Kim, J. Hur, Y.-C. Luo, and S. Yu, "Tunable non-volatile gate-to-source/drain capacitance of FeFET for capacitive synapse," *IEEE Electron Device Letters* 44, no. 10 (2023): 1628–1631.

[87] E. Paasio, *Modelling Ferroelectric HZO-Based Devices for Neuromorphic Computing Applications*, 2023.

[88] E. Paasio, R. Ranta, and S. Majumdar, "A physics-based compact model for ferroelectric capacitors operating down to deep cryogenic temperatures for applications in analog memory and neuromorphic architectures," *Advanced Electronic Materials* 11, no. 9 (2025): 2400840.

[89] S. Majumdar, "Harnessing ferroic ordering in thin film devices for analog memory and neuromorphic computing applications down to deep cryogenic temperatures," *Frontiers in Nanotechnology* 6 (2024): 1371386.

[90] N. Liu, L. Zhao, X. Wang, Y. Cui, and Y. Zhou, "$HfO_2$-based ferroelectric optoelectronic memcapacitors," *IEEE Electron Device Letters* 44, no. 3 (2023): 524–527. https://doi.org/10.1109/LED.2023.3235909.

[91] X. Wang, Y. C. Luo, J. Hur, A. Lu, T. H. Wang, and S. Yu, "Enabling low-power charge-domain nonvolatile computing-in-memory (CIM) with ferroelectric memcapacitor," *IEEE Transactions on Electron Devices* 71, no. 4 (2024): 2404–2410. https://doi.org/10.1109/TED.2024.3367965.

[92] B.-E. Park, H. Ishiwara, M. Okuyama, S. Sakai, and S.-M. Yoon (eds.), Ferroelectric-Gate Field Effect Transistor Memories: Device Physics and Applications, Topics in Applied Physics, vol. 134 (Dordrecht: Springer, 2016). ISBN 978-94-024-0839-3, https://doi.org/10.1007/978-94-024-0841-6.

[93] J. Müller, T. S. Böscke, S. Müller, E. Yurchuk, P. Polakowski, J. Paul, ... and T. Mikolajick, "Ferroelectric hafnium oxide: A CMOS-compatible and highly scalable approach to future ferroelectric memories," in *2013 IEEE International Electron Devices Meeting (IEDM)* (2013): 10.8.1–10.8.4.

[94] K. Kim, M. S. Song, H. Hwang, S. Hwang, and H. Kim, "A comprehensive review of advanced trends: From artificial synapses to neuromorphic systems with consideration of non-ideal effects," *Frontiers in Neuroscience* 18 (2024): 1279708.